\documentclass{article}

\usepackage{multicol}

\usepackage{amsmath,amssymb,amsfonts}   
\usepackage{graphicx,bm}

\newcommand{\e}{{\rm e}}

\newcommand{\x}{{\bf x}}

\renewcommand{\u}{\widetilde{u}}

\newcommand{\tgamma}{\widetilde{\gamma}}

\newcommand{\calG}{{\mathcal G}}
\DeclareMathAlphabet\mathbfcal{OMS}{cmsy}{b}{n}
\renewcommand{\r}{\widetilde{r}}

\begin{document}

\title{Local accumulation times in a diffusion--trapping model of synaptic receptor dynamics}

\author{\em
Ryan D. Schumm and Paul C. Bressloff\\ Department of Mathematics, 
University of Utah \\155 South 1400 East, Salt Lake City, UT 84112}



\maketitle

\begin{abstract} 
The lateral diffusion and trapping of neurotransmitter receptors within the postsynaptic membrane of a neuron plays a key role in determining synaptic strength and plasticity. Trapping is mediated by the reversible binding of receptors to scaffolding proteins (slots) within a synapse. In this paper we introduce a new method for analyzing the transient dynamics of synapses in a diffusion-trapping model of receptor trafficking. Given a population of spatially distributed synapses, each of which has a fixed number of slots, we calculate the rate of relaxation to the steady-state distribution of bound slots (synaptic weights) in terms of a set of local accumulation times. Assuming that the rates of exocytosis and endocytosis are sufficiently slow, we show that the steady-state synaptic weights are independent of each other (purely local). On the other hand, the local accumulation time of a given synapse depends on the number of slots and the spatial location of all the synapses, indicating a form of transient heterosynaptic plasticity. This suggests that local accumulation time measurements could provide useful information regarding the distribution of synaptic weights within a dendrite.
\end{abstract}

\section{Introduction}

The lateral diffusion and trapping of neurotransmitter receptors within the postsynaptic membrane of a neuron plays a key role in mediating synaptic strength and plasticity \cite{Meier01,Borgdorff02,Dahan03,Choquet03,Bredt03,Groc04,Collinridge04,Triller05,Ashby06,Ehlers07,Gerrow10,Henley11,Choquet13,Roth17,Choquet18,Triller19}
The two most studied examples are glycine receptors (GlyRs) and $\alpha$-amino-3-hydroxy-5-methyl-4-isoxazole-propionic acid receptors (AMPARs), although diffusion-trapping appears to be a general mechanism for most types of neurotransmitter receptors. GlyRs are ligand-gated ion channels that mediate chloride-dependent synaptic inhibition, and are found 
 in the postsynaptic membrane of the soma and initial portion of dendrites in spinal cord neurons. 
Single particle tracking (SPT) experiments have shown that
 GlyRs alternate between diffusive and confined states at the cell surface, and confinement is spatially associated with post-synaptic densities (PSDs) that contain the scaffolding protein gephyrin \cite{Meier01,Dahan03}. The majority of fast excitatory synaptic transmission in the central nervous system is mediated by the neurotransmitter glutamate binding to AMPARs located in the postsynaptic membrane of dendritic spines. Again SPT experiments reveal that AMPARs diffuse freely within the dendritic membrane until they
enter a spine, where they are temporarily confined by the geometry of the spine and
through interactions with scaffolding proteins such as PSD-95 and cytoskeletal
elements within the PSD \cite{Borgdorff02,Groc04}. A surface receptor in either type of synapse may also be
internalized via endocytosis and stored within an intracellular compartment, where it is either recycled to the surface via recycling endosomes and exocytosis, or sorted for degradation by late endosomes and lysosomes \cite{Ehlers00}.

A number of models have explored the combined effects of diffusion-trapping and recycling on the number of synaptic AMPARs within dendritic spines \cite{Earnshaw06,Holcman06,Bressloff07,Earnshaw08,Thoumine12,Triesch18}.  In such models, the synapse is treated as a self-organizing compartment in which the number of receptors is a dynamic steady-state that determines the strength of the synapse; activity-dependent changes in the strength of the synapse then correspond to shifts in the dynamical set-point. Most diffusion-trapping models assume that the number of trapping sites or ``slots'' within a given synapse is fixed. However, it is known that scaffolding proteins and other synaptic components are also transported into and out of a synapse, albeit at slower rates \cite{Salvatico15}. Several experimental and modeling studies have analyzed the joint localization of gephyrin scaffolding proteins and GlyRs at synapses \cite{Sekimoto09,Hasel11,Hasel15,Hakim20,Specht21}, showing how stable receptor-scaffold domains could arise dynamically.

In this paper we analyze a diffusion-trapping model of receptor trafficking in the case of a population of spatially distributed synapses, each of which has a fixed number of slots. We use the model to investigate how the spatial locations of the synapses and the distribution of slot proteins affects (i) the steady-state number of bound receptors in each synapse (synaptic weights), and (ii) the approach to steady-state as determined by a set of local accumulation times. The latter provide a new method for characterizing the transient response of synapses. (Accumulation times have previously been used to estimate the time to form a protein concentration gradient during morphogenesis \cite{Berez10,Berez11,Gordon11}.) In particular, we show that for a cluster of closely spaced synapses, the steady-state fraction of bound receptors in each synapse is the same. This implies that an increase in the rate of slot binding under some form of long term potentiation, say, scales all synaptic strengths by the same factor (multiplicative scaling). However, multiplicative scaling breaks down for spatially separated synapses. We then show that, in contrast to the synaptic weights,  the local accumulation times depend on the distribution of slot proteins across all the synapses. For both the steady-state solution and accumulation times, we demonstrate that two different modes of behavior can be observed corresponding to small and large exocytosis rates, respectively.

The structure of the paper is as follows. The basic diffusion-trapping model is introduced in Sect. 2. This takes the form of a quasi one-dimensional (1D) model of a dendritic cable, in which the synapses are represented as point sources or sinks at locations along the cable. We derive a set of reaction diffusion equations for the extrasynaptic receptor concentration and the fraction of bound slot proteins in each synapse. The last quantity is identified with synaptic strength.  In section 3 we show that the steady-state distributions can be obtained by solving the steady-state reaction-diffusion system and expressed in terms of 1D Green's functions of the modified Helmholtz equation along analogous lines to Refs. \cite{Earnshaw06,Bressloff07,Earnshaw08}.  In section 4, a similar procedure is performed to obtain the accumulation times by solving the time-dependent reaction-diffusion equations in Laplace space. 

\setcounter{equation}{0}
\section{Diffusion-trapping model}

\begin{figure}[b!]
\centering
\includegraphics[width=8cm]{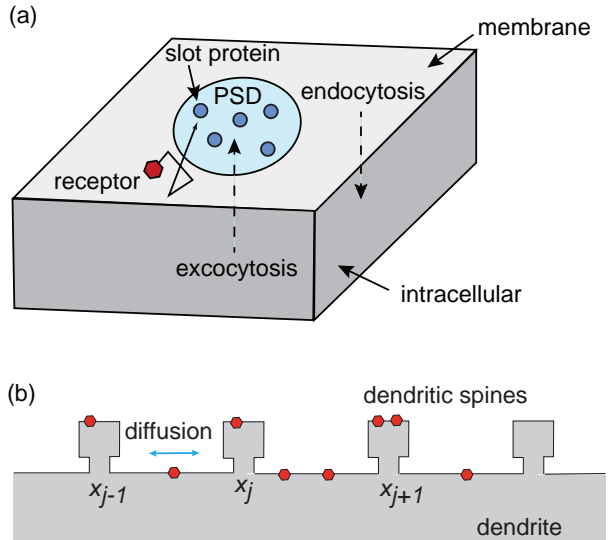}
\caption{(b) Schematic illustration of a diffusion-trapping model for a protein receptor by a single postsynaptic density (PSD). The receptor diffuses freely extrasynaptically, but can transiently bind to scaffolding protein slots within the PSD. The PSD thus acts as a trapping region. Receptors can also be inserted into or removed form the surface via exo/endocytosis. (b) Schematic illustration of the 1D diffusion-trapping model.}
\label{fig1}
\end{figure}

In Fig. \ref{fig1}(a) we show a simplified schematic diagram of a single PSD in the plasma membrane of a neuron. A neurotransmitter receptor diffuses freely in the extrasynaptic membrane, which means that its mean-square displacement (MSD) is proportional to $Dt$, where $t$ is the elapsed time in the free state, and $D$ is the membrane diffusivity. However, on entering the PSD it can reversibly bind to a scaffolding or slot protein, which has the effect of temporarily confining or trapping the receptor within the PSD. This can be amplified by molecular crowding within the PSD, which slows the rate of diffusion, and by geometrical factors such as the narrow neck of a dendritic spine \cite{Holcman04,Biess07,Simon14,Adrian17}. Recent imaging studies of the microstructure of the PSD suggest that there exist PSD microdomains containing higher densities of scaffolding proteins, where most of the receptors are stabilized \cite{MacGillavry11,Nair13}; outside of these domains, receptors tend to diffuse more freely. For simplicity, we ignore the detailed microstructure of a PSD and the geometrical effects of spines (in the case of excitatory synapses) by treating the PSD as a homogeneous medium within which a receptor can bind at a rate that is proportional to the number of slots. We also allow for receptors to be inserted into the surface membrane via exocytosis and internalized via endocytosis \cite{Ehlers00}. Experimentally it has been found that there exist sites of exocytosis and endocytosis within the proximity of synapses as well as various extrasynaptic locations. In our model, we assume that endocytosis occurs everywhere whereas exocytosis occurs (peri) synaptically. However, one could modify these assumptions accordingly. 

Following Refs. \cite{Earnshaw06,Bressloff07,Earnshaw08}, we treat the domain $\Omega$ as a 1D dendritic cable in which the synapses are represented as point sources or sinks on the cable, see Fig. \ref{fig1}(b). This particular approach exploits the quasi 1D geometry of a dendrite over length scales of the order of hundreds of microns, and the fact that dendritic spines act as effective compartments that are joined to the dendrite via narrow (pointlike) necks \cite{Triesch18}.  Additionally, we are leveraging the fact that the Green's function of the 1D diffusion equation is non-singular (in contrast to higher spatial dimensions) so that we avoid introducing singular solutions when modeling the dendritic spines as point--like
sources or sinks \cite{Bressloff08}. 

Suppose that $\Omega$ is a semi-infinite cable with uniform circumference $C$ and axial coordinate $x\in [0,\infty)$; $x=0$ denotes the location of the somatic end of the cable. (Alternatively, we could consider a cable of length $L$ and simply ignore the effects of the distal boundary for sufficiently large $L$.)
Let $x_j$, $j=1,\ldots,N$, denote the position
 of the $j$th spine. The positions are ordered such that $x_j<x_{j+1}$ for all $j=1,\ldots,N-1$. Finally, let $u(x,t)$ denote the concentration of receptors per unit length of cable and
 $S_k$ and $r_k(t)$ be the number of binding sites and fraction of bound sites in synapse $k$ respectively.  This reaction-diffusion system is then defined according to the following system of equations:
\begin{subequations}
\label{trap}
\begin{align}
  \frac{\partial u}{\partial t} &= D\frac{\partial^2 u}{\partial x^2} - \gamma u(x, t)   +\sum_{k=1}^{N}
\left[ S_kg_k(t) + \sigma - \tgamma u(x_k, t) \right]\delta(x - x_k)  ,\\
 \frac{d r_k}{d t} &=  \kappa_-r_k(t) - \kappa_+ u(x_k, t)\left(1 - r_k(t)\right) \equiv -g_k(t)  ,
 \end{align}
\end{subequations}
together with the boundary condition $-D\partial_x u(0,t)=J_0 > 0$. The latter represents a constant flux of receptors injected from the soma. Here $\gamma$ and $\tgamma$ are the rates of extrasynaptic and synaptic endocytosis respectively, $\sigma$ is the rate of exocytosis, $\kappa_{\pm}$ are slot binding/unbinding rates, and $D$ is the extrasynaptic diffusivity.  The point-like synapses are represented by the Dirac delta functions $\delta(x-x_k)$, $k=1,\ldots,N$. Finally, we impose the initial conditions
\begin{equation}
\label{init}
u(x,0)=0, \quad r(0)=0.
\end{equation}

It is important to note that $\gamma,\sigma,\kappa_-$ have units of inverse time, whereas $\widehat{\gamma},\kappa_+$ have units of speed.
Unfortunately, only a few model parameters are known explicitly \cite{Triesch18}. These include the unbinding rate $\kappa_-\sim 2.5\times 10^{-2} s^{-1}$ \cite{Henley16}, the rate of internalization $\gamma  \sim 10^{-3}-10^{-2} s^{-1}$ \cite{Ehlers07} and the membrane diffusivity $D\sim 0.1 \mu m^2s^{-1} $ \cite{Ehlers07}. One general observation, however, is that the basal rates of receptor binding and unbinding are at least an order of magnitude faster than the rates of receptor internalization and externalization.

 In this paper we are interested in calculating the steady-state solution of equations (\ref{trap}) and the corresponding approach to steady-state. We will investigate the latter in terms of  the accumulation times of the reaction-diffusion process. In order to construct the accumulation time of $u(x,t)$ to reach the steady-state $u^*(x)$, consider the function
\begin{equation}
\label{R}
Z(x,t)=1-\frac{u(x,t)}{u^*(x)},
\end{equation}
which represents the fractional deviation of the concentration from the steady-state. Assuming that there is no overshooting, $1-Z(x,t)$ is the fraction of the exterior steady-state concentration that has accumulated at $x$ by time $t$. It follows that $-\partial_t Z(x,t)dt$ is the fraction accumulated in the interval $[t,t+dt]$. The accumulation time is then defined as \cite{Berez10,Berez11,Gordon11}
\begin{equation}
T(x)=\int_0^{\infty} t\left (-\frac{\partial Z(x,t)}{\partial t}\right )dt=\int_0^{\infty} Z(x,t)dt.
\end{equation}
 Finally, by analogy, we can also define an accumulation time for the fraction of bound receptors, namely,
\begin{equation}
\tau_k=\int_0^{\infty} z_k(t)dt,\quad z_k(t)=1-\frac{r_k(t)}{r_k^*}.
\end{equation}

It is often more useful to calculate an accumulation time in Laplace space. First consider $Z(x,t)$ in equation (\ref{R}). Using the identity 
\[u^*(x)=\lim_{t\rightarrow \infty} u(x,t)=\lim_{s\rightarrow 0}s\widetilde{u}(x,s),\]
and setting $\widetilde{F}(x,s)=s\widetilde{u}(x,s)$ 
implies that
\[s\widetilde{Z}(x,s)=1-\frac{\widetilde{F}(x,s)}{\widetilde{F}(x,0)},\]
and
\begin{eqnarray}
T(x)&=&\lim_{s\rightarrow 0} \widetilde{Z}(x,s) = \lim_{s\rightarrow 0}\frac{1}{s}\left [1-\frac{\widetilde{F}(x,s)}{\widetilde{F}(\x,0)}\right ]=-\frac{1}{\widetilde{F}(x,0)}
\left .\frac{d}{ds}\widetilde{F}(x,s)\right |_{s=0}.
\label{acc}
\end{eqnarray}
Similarly, setting $\widetilde{f}_k(s) =s\widetilde{r}_k(s)$, we have
\begin{eqnarray}
\tau_k=-\frac{1}{\widetilde{f}_k(0)}
\left .\frac{d}{ds}\widetilde{f}_k(s)\right |_{s=0}.
\label{acck}
\end{eqnarray}
 
\section{Steady-State Solution} 
At steady state we have $g_k(t)=0$ for all $k=1,\ldots,N$. hence, the steady-state equations take the form
\begin{subequations}
\begin{align}
\label{ssu}
 D\frac{d^2 u^*(x)}{d x^2} -\gamma u^*(x)&=-\sum_{k=1}^{N}
[ \sigma-\widehat{\gamma} u^*(x_k)]\delta(x-x_k) ,\\
r_k^*&=\frac{\kappa_+u^*(x_k)}{\kappa_-+\kappa_+u^*(x_k)},
\end{align}
\end{subequations}
and $du^*/dx=-J_0 / D$ at $x=0$.
Equation (\ref{ssu}) can be solved in terms of the
1D Neumann Green's function $G(x,\xi)$ on $[0,\infty)$, which is the solution to the equation
\begin{equation}
\label{NG1D0}
D\frac{d^2 G(x,\xi)}{d x^2}-\gamma G(x,\xi)=-\delta(x-\xi),\quad \left . \frac{d G(x,\xi)}{d x}\right |_{x=0}=0,
 \end{equation}
with $G(x,\xi)\rightarrow 0$ as $|x|\rightarrow \infty$.
One finds that
\begin{align}
\label{NG1D}
&G(x,\xi)=\frac{1}{2\sqrt{D \gamma}}\left [\e^{-\sqrt{ \gamma/D}|x-\xi|}+\e^{-\sqrt{ \gamma/D}(x+\xi)}\right ].
\end{align}
It follows that the
dendritic surface receptor concentration has an implicit solution of
the form
\begin{equation}
  \label{eq:Uchi}
  u^*(x)=J_0G(x,0)+\sum_{k=1}^N [\sigma-\widehat{\gamma} u^*(x_k)]G(x,x_k).
\end{equation}

We can now generate a closed set of equations for the synaptic concentrations $u_j^*=u^*(x_j)$, $j=1,\ldots,N$, by setting
$x=x_j$ in equation (\ref{eq:Uchi}):
\begin{align}
  u_j^*&= J_0G(x_j,0)+ \sum_{k=1}^N G(x_j,x_k)[\sigma -\widehat{\gamma}u^*_k].
    \label{eq:Ui}
\end{align}
This can be rewritten as the matrix equation
\begin{subequations}
\label{matrix-ss}
\begin{equation}
\sum_{k=1}^N[\delta_{j,k}+\frac{\widehat{\gamma}}{\sqrt{\gamma D}} {\bm \calG}_{jk}]u_k^*= \textbf{H}_j
\end{equation}
with
\begin{align}
{\bm \calG}_{jk}&= \sqrt{\gamma D}G(x_j,x_k),\quad
 \textbf{H}_j=J_0G(x_j,0)+ \sigma\sum_{k=1}^N  G(x_j, x_k).
 \end{align}
 \end{subequations}
 A solution for $u_k^*$, $k=1,\ldots,N$ will exist provided that the matrix ${\bm I} + \epsilon\gamma_0{\bm {\mathcal G}}$ is invertible. One way to invert the matrix is to exploit the fact that the rates of endocytosis are relatively small. That is, since $G(x,x') \sim 1/\sqrt{D\gamma}$, we take
 \begin{equation} 
 \label{approx}
 \frac{\widehat{\gamma}}{\sqrt{D\gamma}}=\epsilon \gamma_0 \ll 1,\end{equation} 
 where $\gamma_0$ is a dimensionless $O(1)$ parameter.
We can now carry out a perturbation expansion of the steady-state solution (\ref{eq:Ui}) with respect to $\epsilon$. That is, write
\begin{equation}
({\bf I}+\epsilon \gamma_0 {\bm \calG})^{-1}={\bf I}-\epsilon \gamma_0 {\bm \calG}+\epsilon^2\gamma_0^2{\bm \calG}^2+\ldots,
\end{equation}
and introduce the series expansion
\begin{equation}
u_j^*= {u}_{j,0}^*+\epsilon u_{j,1}^*+\epsilon^2 u_{j,2}^*+\ldots.
\end{equation}
Substituting into equation (\ref{eq:Ui}) gives
\begin{align}
&{u}_{j,0}^*+\epsilon u_{j,1}^*+\epsilon^2 u_{j,2}^*+\ldots  =\sum_{k=1}^N\left [\delta_{j,k}-\epsilon \gamma_0{\bm \calG}_{jk}+\epsilon^2\gamma_0^2[{\bm \calG}^2]_{jk}+\ldots\right ]\nonumber \\
&\hspace{5cm} \times \left [J_0G(x_k,0)+   \sigma\sum_{l=1}^N  G(x_k,x_l)\right ].
\end{align}
Hence, collecting terms in powers of $\epsilon$ leads to a hierarchy of equations, the first few of which are as follows:
\begin{subequations}
\label{hier}
\begin{align}
{u}_{j,0}^* &=J_0G(x_j,0)+\sigma\sum_{l=1}^N  G(x_j,x_l),\\
u_{j,1}^*&= -\gamma_0\sum_{k=1}^N{\bm \calG}_{jk}u^*_{k,0}.
\end{align}
\end{subequations}
Note that the leading order term ${u}_{j,0}^*$ has two distinct contributions. The first, $J_0G(x_j,0)$, represents the exponential-like decay of the flux strength with respect to the spatial distance of the $j$th synapse from the soma. The second, $\sigma\sum_{l=1}^N  G(x_j,x_l)$, is a pair-wise synaptic interaction term. Clearly, the former contribution will dominate at proximal locations and relatively large flux amplitudes $J_0$. On the other hand, the latter contribution will dominate at more distal locations, particularly when there are a large number of synapses in a cluster. Finally, substituting the series expansion of $u_j^*$ into equation (\ref{eq:Uchi}) shows that $u^*(x)=u_0^*(x)+O(\epsilon)$ with
\begin{equation}
  \label{eq:Uchi0}
  u_0^*(x)=J_0G(x,0)+\sigma \sum_{k=1}^N G(x,x_k).
\end{equation}
Note that $u_{j,0}^*=u_0^*(x_j)$.

\begin{figure*}[b!]
\centering
\includegraphics[width=10cm]{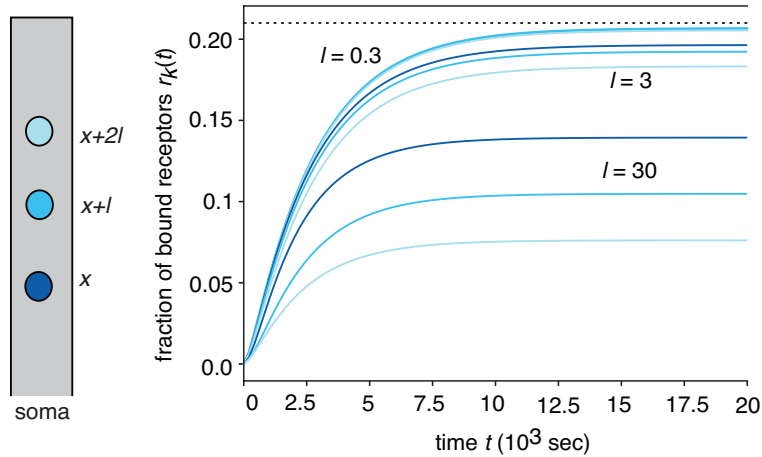}
\caption{Results of numerically simulating equations (\ref{trap}) for a cluster of three identical synapses at positions $x_1=x,x_2=x+\ell$ and $x_3=x+2\ell$. Plots of $r_k(t)$, $k=1,2,3$, for various synaptic spacings $\ell$. The group of synapses starts at $x=5$ $\mu$m. In the tightly clustered case ($\ell=0.3$ $\mu$m), all three curves lie on top of each other so only one is shown. The dashed horizontal line indicates the steady-state fraction $R=0.21$. The other parameter values are $\sigma=10^{-3}$ s$^{-1}$, $\gamma=10^{-3}$ s$^{-1}$, $\widehat{\gamma}=5\times 10^{-4}$ $\mu$m/s, $D=0.1$ $\mu$m$^2$/s, $S_k = 10$, $\kappa_+ = 10^{-3}$ $\mu$m/s, $\kappa_- = 10^{-3}$ s$^{-1}$, and $J_0 = 10^{-3}$ s$^{-1}$.}
\label{fig2}
\end{figure*}

\begin{figure*}[t!]
\centering
\includegraphics[width=12cm]{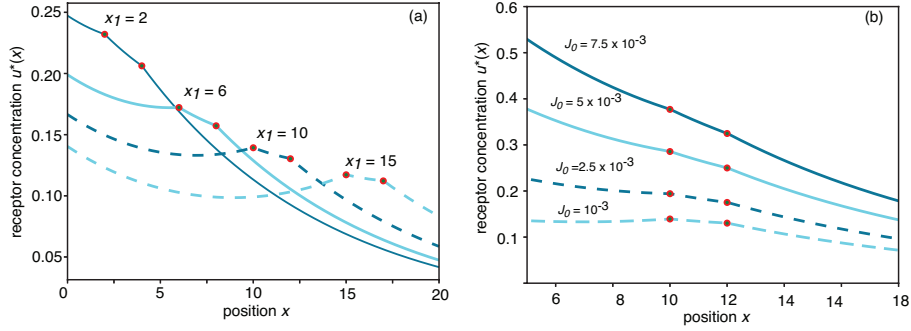}
\caption{Steady-state receptor concentration $u^*(x)$ for a pair of synapses separated by a fixed distance $\ell=2$ $\mu$m. Other parameter values are as in Fig. \ref{fig2}. The concentration $u^*(x)$ is obtained by numerically inverting the matrix equation (\ref{matrix-ss}) for $u_k^*$ and substituting the result into equation (\ref{eq:Uchi}). (a) Plots of $u^*(x)$ as a function of $x$ for different spatial locations of the synaptic pair and fixed $\ell$. (b) Corresponding plots for different flux values $J_0$, and fixed synaptic locations $x_1 = 10$ $\mu$m, $x_2 = 12$ $\mu$m.
}
\label{fig3}
\end{figure*}

\begin{figure*}[b!]
\centering
\includegraphics[width=8cm]{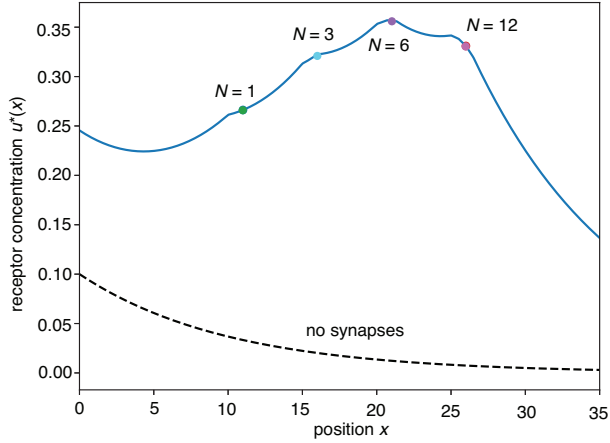}
\caption{Plot of the steady-state concentration $u^*(x)$ (obtained along identical lines to Fig. \ref{fig3}) in the case of four synaptic clusters at $x=10$ $\mu$m, $x=15$ $\mu$m, $x=20$ $\mu$m, and $x=25$ $\mu$m, respectively. Here $\sigma=0.01$ s$^{-1}$ and other parameters are as in Fig. \ref{fig2}. The synapses in each cluster are distributed over a domain of size $\ell=2$ $\mu$m and the number of synapses in each cluster is varied. The dashed curve corresponds to the $O(1)$ bulk concentration when $\sigma=0$, which is the same as when there are no synapses.}
\label{fig4}
\end{figure*}

If the synapses are clustered in an interval of length $[X,X+\ell]$ with $\ell \ll \sqrt{D/\gamma}$, then $G(x_j,0)\approx G(X,0)$ and $G(x_k,x_l)\approx G(X,X)$ so that
\begin{equation}
u_{j,0}^*\approx U(X)\equiv J_0G(X,0)+\sigma N G(X,X),
\end{equation}
and
\begin{equation}
\label{R_frac}
r_j^*\approx R(X) \equiv \frac{\kappa_+U(X)}{\kappa_-+\kappa_+U(X)} 
\end{equation}
for all $j=1,\ldots,N$, where $N$ is the number of synapses in the cluster.
 We thus find that the fraction of filled slots in each synapse is the same, irrespective of the size of the synapse.
Suppose that $w_j=S_j R$ is identified as the strength of the $j$th synapse. This implies that the ratio of synaptic strengths of any two synapses within the cluster is equal to the ratio of their number of slots. Moreover, if $R$ is modified due to changes in the binding rate $\kappa_+$, say, then all synaptic strengths are scaled by the same factor (multiplicative scaling). An analogous result was previously obtained in a non-spatial compartmental model \cite{Triesch18}.
On the other hand, as synapses within a cluster become more separated, the multiplicative scaling rule breaks down. This is illustrated in Fig. \ref{fig2}, where we show results from numerically simulating the full model (\ref{trap}) for a cluster of three synapses with nearest-neighbor distance $\ell$. It can be seen that the fraction of bound receptors $r_k(t)$, $k=1,2,3$, converges to a steady state $r_k^*$ as $t\rightarrow \infty$. For sufficiently small $\ell$ the time-dependent curves $r_k(t)$ are approximately independent of $k$ and equation (\ref{R_frac}) provides a good approximation of the steady-state value. However, as the separation between the synapses increases, these curves asymptote to different steady-state values that are all lower than the original steady-state. Therefore, tightly clustering the synapses has the effect of both homogenizing and increasing the steady-state synaptic strengths. The multiplicative scaling of the synaptic strengths also breaks down when the effect of the somatic receptor flux becomes sufficiency strong.  This can be achieved by either moving a synapse cluster closer to $x=0$ or by increasing $J_0$, as illustrated in Fig. \ref{fig3}.  Finally, we note that if $\sigma >0$ then the strengths of the synapses in a cluster are highly dependent on the number of synapses in the cluster. Figure \ref{fig4} shows that the synaptic strength significantly increases with the number of synapses in a cluster even if that cluster is moved farther from the receptor flux at $x=0$.

\section{Local accumulation times} 

Having obtained the steady-state solution, we can now calculate the accumulation times using equations (\ref{acc}) and (\ref{acck}). In order to Laplace transform equations (\ref{trap}), we will assume that concentration of receptors within a synapse is sufficiently small so that $\kappa_-\gg \kappa_+u(x_k,t)$  and $r_k(t)\ll 1$. This means that nonlinear effects due to the saturation of the binding sites can be ignored. 
Laplace transforming equations (\ref{trap}a-c) using the
initial conditions $u(x,0)=0$ and $r_j(0)=0$, then gives
\begin{subequations}
\label{RDLT}
\begin{align}
& D\frac{\partial^2 \widetilde{u}(x,s)}{\partial x^2}-(s+\gamma)\widetilde{u}(x,s) \nonumber
 \\ \quad  &=- \sum_{j=1}^{N}\bigg (S_j [\kappa_-\widetilde{r}_j(s)-\kappa_+\widetilde{u}(x_j,s)]+\frac{\sigma}{s} -\widehat{\gamma}\widetilde{u}(x_j,s)\bigg ) \delta(x-x_j),\\
\widetilde{r}_k(s)&=\frac{\kappa_+}{\kappa_-+s} \widetilde{u}(x_k,s).
\end{align}
\end{subequations}
Analogous to equation (\ref{NG1D0}), we introduce the $s$-dependent Neumann Green's function on $[0,\infty)$ according to
\begin{equation}
\label{GG}
D\frac{\partial^2 G(x,y;s)}{\partial x^2}-(s+\gamma)G(x,y;s)=-\delta(x-y),\quad \left . \frac{\partial G(x,y;s)}{\partial x}\right |_{x=0}=0.\end{equation}
 From equation (\ref{NG1D}) we have
\begin{align}
G(x,y;s)&=\frac{1}{2\sqrt{D[s+\gamma]}}\left [\e^{-\sqrt{(s+\gamma)/D}|x-y|}+\e^{-\sqrt{(s+\gamma)/D}(x+y)}\right ].
\label{Gs}\end{align}
It then follows that
\begin{align}
\label{tilu}
&\widetilde{u}(x,s)=\frac{J_0}{s} G(x,0;s)+\sum_{k=1}^{N}\bigg (
\frac{\sigma}{s}-\left [ \frac{sS_k\kappa_+}{\kappa_-+s}+\widehat{\gamma} \right ] \widetilde{u}(x_k,s) \bigg )G(x,x_k,s).
\end{align}
Multiplying this equation by $s$ and taking the limit $s\rightarrow 0$ immediately recovers the steady-state solution (\ref{eq:Uchi}), since $G(x,x_k,0)=G(x,x_k)$.

It remains to determine the synaptic terms $\widetilde{u}(x_j,s)$. Setting $x=x_j$ in equation (\ref{tilu}) gives the matrix equations
\begin{align}
  \label{eq:Uis}
  &\u(x_j,s) = \frac{J_0}{s}G(x_j,0,s) +\sum_{k=1}^{N}\bigg (
\frac{\sigma}{s}-\left [ \frac{sS_k\kappa_+}{\kappa_-+s}+\widehat{\gamma} \right ] \widetilde{u}(x_k,s) \bigg )G(x_j,x_k,s).
\end{align}
This can be rewritten as the matrix equation
\begin{align}
&\sum_{k=1}^N\left (\delta_{j,k}+\left [s \Gamma(s)S_k+\frac{\widehat{\gamma}}{\sqrt{[s+\gamma]D}} \right ] {\mathcal G}_{jk}(s)\right ) \u(x_k,s)= \frac{H_j(s)}{s},
\label{Us}
\end{align}
with
\begin{align}
{\mathcal G}_{jk}(s)&= \sqrt{[s+\gamma]D}G(x_j,x_k,s),\\
H_j(s)&=J_0G(x_j,0,s)+ \sigma \sum_{k=1}^N  G(x_j,x_k,s),\\
 \Gamma(s)&=\frac{\kappa_+}{\kappa_-+s}\frac{1}{\sqrt{[s+\gamma]D}}.
 \end{align}
 Note that $H_j(0)=u_{0}^*(x_j)$ with $u_0^*(x)$ defined by equation (\ref{eq:Uchi0}).
 Again we will  invert the matrix by exploiting the relatively slow rate of endocytosis. Since we are ultimately interested in the limit $s\rightarrow 0$ we will also assume that $s$ is small and impose the inequalities (\ref{approx}) with $s=O(\epsilon)$.
We can now carry out a perturbation expansion of equation (\ref{Us}) with respect to $\epsilon$ by writing
\begin{align}
&({\bf I}+{\bm \calG}(s)[s\Gamma(s){\bf S}+\epsilon \gamma_0{\bf I}] )^{-1}={\bf I}- {\bm \calG}(s)[s\Gamma(s){\bf S}+\epsilon \gamma_0{\bf I}] +\ldots,
\end{align}
with ${\bf S}=\mbox{diag}(S_1,S_2,\ldots,S_N)$,
and introduce the series expansion
\begin{align}
\u(x_k,s)&=\u_0(x_k,s)+\epsilon \u_1(x_k,s)+\epsilon^2 \u_2(x_k,s) +\ldots.
\end{align}
Substituting into equation (\ref{eq:Uis}) gives
\begin{align}
&\u_0(x_j,s)+\epsilon \u_1(x_j,s)+\ldots  =\frac{1}{s}\sum_{k=1}^N\left [\delta_{j,k}-{ \calG}_{jk}(s)(s\Gamma(s)S_k+\epsilon \gamma_0)  +\ldots\right ]H_k(s).
\end{align}
We thus obtain the leading order solution
\begin{align}
\label{hieracc}
\u_0(x_j,s)&=\frac{1}{s}\left [H_j(s)-s\Gamma(s)\sum_{k=1}^N\calG_{jk}(s)S_kH_k(s)\right ].
\end{align}

Having obtained a series expansion of $\u(x_j,s)$ we have a corresponding series expansion of the fraction of bound receptors according to equation (\ref{RDLT}c):
\begin{align}
\r_j(s)&=\r_{j,0}(s)+\epsilon \r_{j,1}(s)+\ldots,\\
 \r_{j,n}(s)&=\frac{\kappa_+}{\kappa_-+s} \widetilde{u}_n(x_j,s).
\end{align}
Substituting into equation (\ref{acck}) yields
\begin{align}
\tau_j&=-\frac{1}{r_j^*}
\left .\frac{d}{ds}\bigg [\frac{s\kappa_+}{\kappa_-+s}\left (\u_{0}(x_j,s)+\epsilon \u_{1}(x_j,s)+\ldots\right )\bigg ]\right |_{s=0}\nonumber \\
& =-\frac{1}{r_{j,0}^*}\frac{d}{ds}\bigg [ \frac{\kappa_+}{\kappa_- + s} \bigg (H_j(s)\left . -s\Gamma(s)\sum_{k=1}^N\calG_{jk}(s)H_k(s)\bigg)\bigg ]\right |_{s=0}+O(\epsilon)\nonumber \\
&=-\frac{1}{r_{j,0}^*}\frac{\kappa_+}{\kappa_-}\bigg [  \partial_s H_j(0) - \frac{1}{\kappa_-}H_j(0)  -\Gamma(0)\sum_{k=1}^N\calG_{jk}(0)S_kH_k(0))\bigg ] +O(\epsilon).
\end{align}
In the unsaturated regime,
\begin{equation}
r_{j,0}^*=\frac{\kappa_+H_j(0)}{\kappa_-}.
\end{equation}
Hence, to leading order, we have
\begin{align}
\label{accum-time-rec}
\tau_j= \frac{|\partial_sH_j(0)|}{H_j(0)}+\frac{\kappa_+}{\kappa_- }\sum_{k=1}^N S_k \frac{H_k(0)}{H_j(0)} G(x_j, x_k) + \frac{1}{\kappa_-} .
\end{align}
In the limit $\sigma\rightarrow 0$, we obtain the simplified expression
\begin{align}
\label{accum-time-rec2}
\tau_j= T_0(x_j)+\frac{\kappa_+}{\kappa_- }\sum_{k=1}^N S_k \frac{G(x_k,0)}{G(x_j, 0)} G(x_j, x_k) + \frac{1}{\kappa_-} ,
\end{align}
where
\begin{align}
T_0(x)&=\frac{1}{2}\left [\frac{1}{\gamma}+\frac{x}{\sqrt{D\gamma}} \right ]
\end{align}
is the accumulation time without synapses \cite{Berez10,Berez11,Gordon11}. We have used the result
\begin{align*}
\partial_s G(x,y;s) &=-\frac{1}{4\sqrt{D[s+\gamma]}}\left [\frac{1}{\gamma+s}+\frac{|x-y|}{\sqrt{D[\gamma+s]}} \right ]\e^{-|x-y|\sqrt{(s+\gamma)/D}}\\
&\quad -\frac{1}{4\sqrt{D[s+\gamma]}}\left [\frac{1}{\gamma+s}+\frac{x+y}{\sqrt{D[\gamma+s]}} \right ]\e^{-(x+y)\sqrt{(s+\gamma)/D}},
\end{align*}
which implies 
\begin{align*}
\partial_s G(x,0;s) &=-\frac{1}{2\sqrt{D[s+\gamma]}}\left [\frac{1}{\gamma+s}+\frac{x}{\sqrt{D[\gamma+s]}} \right ]\e^{-x\sqrt{(s+\gamma)/D}}.
\end{align*}
Substituting equations (\ref{tilu}) and (\ref{hieracc}) into equation (\ref{acc}) yields the following equation for the accumulation time in the bulk
\begin{align}
\label{accum-time-bulk}
T(x) = \frac{J_0|\partial_sG(x, 0; 0)|}{u_0^*(x)} + \frac{\kappa_+}{\kappa_- }\sum_{k=1}^NS_k \frac{u_0^*(x_k)}{u_0^*(x)}G(x, x_k) + O(\epsilon) .
\end{align}
In the limit that $\sigma \to 0$, we obtain
\begin{align}
\label{accum-time-bulk2}
T(x) = T_0(x) + \frac{\kappa_+}{\kappa_- }\sum_{k=1}^NS_k \frac{G(x_k, 0)}{G(x, 0)}G(x, x_k) + O(\epsilon) .
\end{align}

\begin{figure*}[t!]
\centering
\includegraphics[width=12cm]{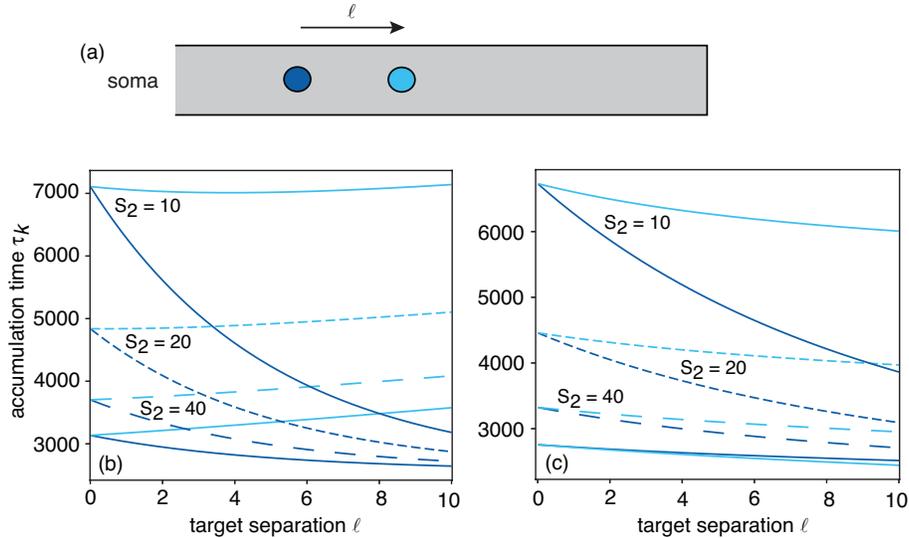}
\caption{Pair of synapses with $x_1=10$ $\mu$m and $x_2=x_1+\ell$. (a) Plot of the accumulation times $\tau_1$ (dark curves) and $\tau_2$ (light curves) as a function of the separation $\ell$ for $\sigma=0$, $S_1=10$ and various values of the slot protein number $S_2$: $S_2=10,20,40,80$. (b) Corresponding plots for $\sigma=0.1$. The other parameter values are as in Fig. \ref{fig2}. The accumulation times $\tau_{1,2}$ are calculated using equation (\ref{accum-time-rec}).}
\label{fig5}  
\end{figure*}

\begin{figure*}[t!]
\centering
\includegraphics[width=12cm]{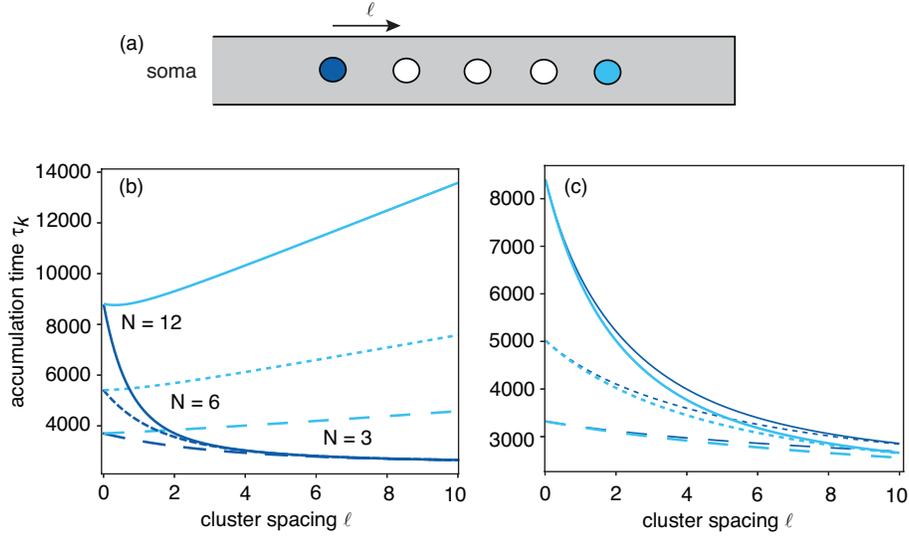}
\caption{Cluster of $N$ synapses at spatial locations $x_k = x_1 + (k - 1)\ell$ where $x_1 = 10$ $\mu$m. (a) Plots of the accumulation times $\tau_1$ and $\tau_N$, respectively, for the first and last synapses in the cluster as a function of the cluster spacing $\ell$ and various cluster sizes $N=3,6,12$ for $\sigma=0$ . (b) Corresponding plots for $\sigma=0.1$. The other parameter values are as in Fig. \ref{fig2}. The accumulation times $\tau_{1,N}$ are calculated using equation (\ref{accum-time-rec}).} 
\label{fig6}
\end{figure*}

\begin{figure*}[t!]
\centering
\includegraphics[width=12cm]{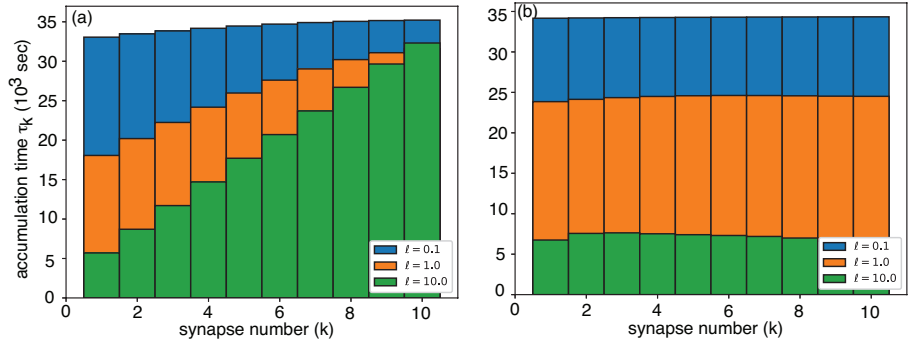}
\caption{Dependence on synaptic spacing $\ell$ for a cluster of $N=10$ identical synapses with $S_k=50$. (a) Plots of the accumulation times $\tau_k$ given by equation (\ref{accum-time-rec2}) in the case $\sigma=0$. The synapses are located at the positions $x_k=x_1+(k-1)\ell$ with $x_1=10$ $\mu$m and $\ell$ varied. (b) Corresponding plots for $\sigma=10^{-3}$. The other parameter values are as in Fig. \ref{fig2}.} 
\label{fig7}
\end{figure*}

Comparison of equations (\ref{accum-time-rec}) and (\ref{accum-time-rec2}) with the steady-state solution (\ref{hier}) establishes a number of significant differences. First, in the limit $\sigma \rightarrow 0$ (no exocytosis), the $O(1)$ steady-state receptor concentration is purely local, $u_{j,0}^*=J_0G(x_j,0)$. This means that the presence of a set of synapses does not affect the $O(1)$ bulk concentration $u_0^*(x)$. Hence, the fraction of bound receptors in a given synapse is independent of the other synapses. On the other hand, synaptic interactions have an $O(1)$ effect on the local accumulation time $\tau_j$, even when $\sigma =0$. Second, this $O(1)$ contribution depends on the number of slot proteins $S_k$, $k=1,\ldots,N$, in all the synapses. These features are illustrated in Fig. \ref{fig5}, where we plot the accumulation times $\tau_j$, $j=1,2$ for a pair of synapses with $x_1$ fixed and $x_2=x_1+\ell$ for a variable spacing $\ell$. The dependence of $\tau_j$ on $\ell$ and $ S_2$ can be seen even when $\sigma=0$. The accumulation time $\tau_2$ has a weaker dependence on $\ell$ than $\tau_1$, while the difference $\tau_2-\tau_1$ for fixed $\ell$ is reduced as $\sigma$ increases from zero. In Fig. \ref{fig6} we show analogous plots for $\tau_1$ and $\tau_N$ in a cluster of $N$ synapses with variable spacing $\ell$. It can be seen that adding synapses to a highly separated cluster does not effect the value of $\tau_1$ but increases $\tau_N$ significantly due to the other synapses inhibiting access to the receptor pool.
In the case of a tight cluster of synapses, the accumulation time is approximately the same for each synapse in the cluster. As the spacing $\ell$ is increased, the synapses located closer to $x=0$ reach steady-state more rapidly while the accumulation times of the synapses on the right end of the domain do not change significantly, see Fig. \ref{fig7}(a).  On the other hand, if $\sigma\geq J_0$ and $S_i=S$ for all $j$, then the homogeneity of the synaptic accumulation times is preserved even when the synapses have a large separation.  We also observe that the accumulation times are reduced significantly for all synapses in the cluster, see Fig. \ref{fig7}(b).

\begin{figure*}[b!]
\centering
\includegraphics[width=12cm]{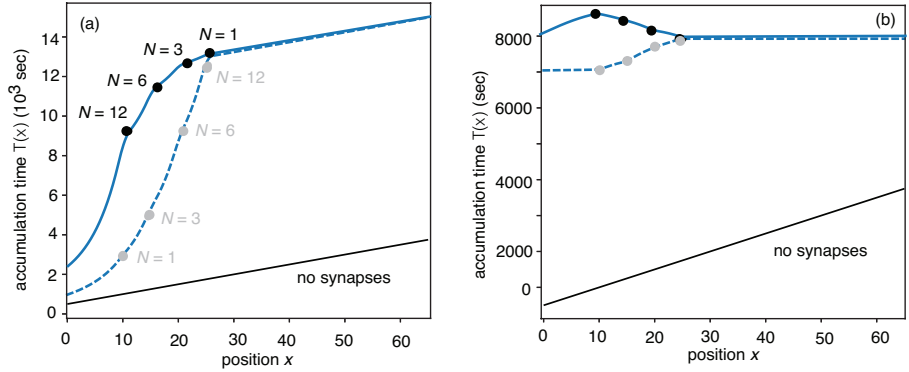}
\caption{(a) Plots of the bulk accumulation time $T(x)$ given by equation (\ref{accum-time-bulk2}) for $\sigma=0$ in the case of four synaptic clusters at $x=10$ $\mu$m, $x=15$ $\mu$m, $x=20$ $\mu$m, and $x=25$ $\mu$m, respectively. The synapses in each cluster are distributed over a domain of size $\ell=2$ $\mu$m and the number of synapses in each cluster is varied. (b) Corresponding plots for $\sigma=0.1$. Other parameter values are as in Fig. \ref{fig2}.} 
\label{fig8}
\end{figure*}

$O(1)$ synaptic interactions also appear in the bulk accumulation time $T(x)$. This is illustrated in Fig. \ref{fig8}, where we show example plots of $T(x)$ as a function of dendritic location $x$. We assume that there are several synaptic clusters of different sizes. In the absence of any synapses, the corresponding accumulation time $T_0(x)$ increases
linearly with $x$. On the other hand, the presence of synaptic clusters significantly increases $T(x)$, which now varies nonlinearly with $x$. If $\sigma\ll J_0$, then $T(x)$ is a monotonically increasing function of $x$ with $T'(x)$ having a local maximum in a neighborhood of each cluster, see Fig. \ref{fig8}(a). If $\sigma \geq J_0$ then $T(x)$ is no longer monotonic and shows a much weaker dependence on $x$ as demonstrated in Fig. \ref{fig8}(b). These two distinct modes of behavior seem to arise from the relative dominance of the two sources of receptors.  When the local pool of receptors available to a cluster is dominated by exocytosis, the accumulation time becomes homogeneous and is mostly a function of the total number of synapses in each cluster.  When the local pool of receptors is dominated by the somatic flux at $x=0$, the heterogeneity of the accumulation times increases and they become mostly a function of their position relative to $x=0$.

\begin{figure*}[t!]
\centering
\includegraphics[width=12cm]{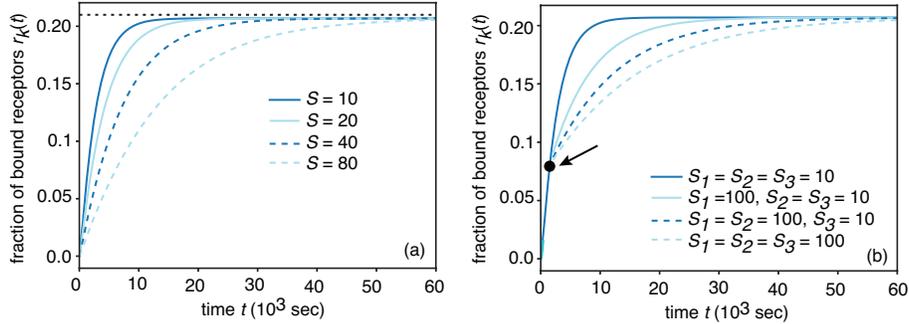}
\caption{Transient heterosynaptic effects in a cluster of three synapses located at $x_k = x + (k - 1)\ell$ with $x=5$ $\mu$m and $\ell=0.1$ $\mu$m. (a) Plots of the fraction of bound receptors $r_k(t)$ for $S_k=S$, $k=1,2,3$, and various values of $S$. (b) Corresponding plots when the number of slot proteins in one or more of the synapses is increased to $S_k=100$ at $t=1500$ $s$ (as indicated by the arrow). The other parameter values are as in Fig. \ref{fig2}. Results are obtained by numerically simulating the full equations (\ref{trap}).}  
\label{fig9} 
\end{figure*}

Finally, the transient heterosynaptic effects of the number of slot proteins $S_k$, $k=1,\ldots,N$, across a cluster of synapses is illustrated in Fig. \ref{fig9}. We see that the steady-state values of the fractions of bound slot proteins is independent of $S_k$ but the total time to steady-state increases with $S_k$. This is consistent with the accumulation time calculations.  Furthermore, increasing the number of slot proteins in a subset of the synapses in a cluster before the steady-state value is reached has the effect of increasing the time to steady-state for all the synapses in the cluster including the unaltered synapses. This implies that the total number of slot proteins in a cluster is the important parameter that controls the dynamics of a synapse rather than the number of slot proteins in an individual synapse.

\section{Discusion}
In this paper, we used a diffusion-trapping model of dendritic receptor trafficking to investigate the effects of spatial heterogeneity on the steady-state distribution of synaptic weights and the relaxation to steady-state. The model took the form of a set of reaction-diffusion equations, in which the dendrite was represented as a semi-infinite cable and each synapse was taken to be a point source. The latter leveraged the non-singular behavior of one-dimensional Green's functions. 
We first derived a solution to the associated steady-state equations, which took the form of a perturbation series expansion in the small dimensionless parameter $\epsilon$ that characterized the relatively slow rates of endocytosis. We showed that for a cluster of closely spaced synapses, the steady-state fraction of bound receptors in each synapse is the same to leading order (multiplicative scaling). However, it was found that the synaptic strengths are coupled at $O(\epsilon)$, and a break down of multiplicative scaling can be observed by increasing the size of the synaptic domain or by increasing the strength of the source of receptors from the soma.  The later was achieved by both moving the cluster closer to $x=0$ or increasing the receptor flux.  Additionally, we observed that the strengths of all synapses in a cluster can be increased by reducing the size of the synaptic domain or increasing the number of synapses in the cluster.

In order to study the temporal dynamics of the receptor trafficking process, we calculated a set of local accumulation times. In contrast to spectral methods, which provide a global measure of the rate at which the system reaches steady-state, the accumulation times determine how rapidly steady-state is approached at each point in the bulk and at each synapse.  The accumulation times were calculated by solving the linearized reaction-diffusion equations in Laplace space using Green's function methods analogous to those used to obtain the steady-state solution. 
Contrary to the results obtained from the steady-state analysis, we found that the accumulation times had a strong dependence on the number of slot proteins in the synapses at order $O(1)$. We thus observed that dynamically altering the number of slot proteins in a single synapse or subset of synapses within a cluster altered the trajectories to steady-state as a whole rather than 
altering the trajectories of only the modified synapses. This implies that the total number of slot proteins in a cluster is the important parameter for controlling the dynamics of the trajectories rather than the distribution of slot proteins among the individual synapses within a cluster. 

Interestingly, two distinct modes of behavior were observed, depending on whether exocytosis or the somatic flux was the dominant source of receptors.
The $O(1)$ dependence on the spatial distribution of synapses within a cluster was similar to the steady-state results when $\sigma > 0$ but this dependence could be eliminated such that the synaptic accumulation times were homogeneous even for large synaptic separations by letting $\sigma\to 0$.  Furthermore, when $\sigma$ was sufficiently large, the presence of a cluster of synapses not only caused a significant increase in accumulation times of the points near the cluster, this increase was communicated to all points to the right of the cluster as well. 

One natural extension of the current model is to consider diffusion in a two-dimensional (2D) cellular membrane, rather than a 1D dendritic cable in which the synapses are represented as point sources or sinks at locations $x_k$ on the cable. The latter exploits the quasi-1D geometry of a dendrite over length scales of the order of hundreds of microns. However, the quasi-1D approximation is not appropriate for synapses distributed over a more local region of a dendrite nor for synapses located in the somatic membrane. In such cases one has to take $\Omega$ to be a 2D domain. Moreover, one can no longer treat the synapses as point-like, since the corresponding 2D Green's function has a logarithmic singularity. However, if the synapses are relatively small compared to the size of the domain, then one can use asymptotic perturbation methods along the lines of \cite{Bressloff08}.


\vskip6pt

\enlargethispage{20pt}








\end{document}